\documentclass[superscriptaddress,twocolumn,showpacs,aps,prl]{revtex4}

\usepackage{amssymb,xcolor}
\usepackage{amsmath}
\usepackage{amsbsy}
\usepackage{amsthm}
\usepackage{graphicx}
\usepackage{amsfonts}
\usepackage{epstopdf}
\usepackage{times}
\usepackage{txfonts}

\newcommand{\beq}{\begin{equation}}
\newcommand{\eeq}{\end{equation}}

\newcommand{\bra}[1]{\langle#1|}

\newcommand{\ket}[1]{|#1\rangle}

\newcommand{\id}{\leavevmode\hbox{\small1\normalsize\kern-.33em1}}

\newcommand{\eq}[1]{Eq. (\ref{#1})}

\newcommand{\qd}{${\cal D}$}

\begin{document}

\title{Extremal Quantum Correlations: Experimental Study with Two-qubit States}

\author{A. Chiuri}
\affiliation{Dipartimento di Fisica, Sapienza {Universit\`a} di Roma, Piazzale Aldo Moro 5, I-00185 Roma, Italy}
\author{G. Vallone}
\affiliation{Museo Storico della Fisica e Centro Studi e Ricerche Enrico Fermi, Via Panisperna 89/A, Compendio del Viminale, I-00184 Roma, Italy}
\affiliation{Dipartimento di Fisica, Sapienza {Universit\`a} di Roma, Piazzale Aldo Moro 5, I-00185 Roma, Italy}
\author{M. Paternostro}
\affiliation{School of Mathematics and Physics, Queen's University, Belfast BT7 1NN, United Kingdom}
\author{P. Mataloni}
\affiliation{Dipartimento di Fisica, Sapienza {Universit\`a} di Roma, Piazzale Aldo Moro 5, I-00185 Roma, Italy}
\affiliation{Istituto Nazionale di Ottica (INO-CNR), L.go E. Fermi 6, I-50125 Firenze, Italy}

\date{\today}

\begin{abstract}
We explore experimentally the space of two-qubit quantum correlated mixed states, including 
frontier ones as defined by the use of quantum discord and von Neumann entropy. Our 
experimental setup is flexible enough to allow for the high-quality generation of a vast variety of states.
We address quantitatively the relation between quantum discord and a recently suggested alternative measure of quantum correlations.
 
\end{abstract}


\maketitle


Entanglement, {\it ``the characteristic trait of quantum mechanics"} according to the words of E. Schr\"odinger~\cite{Sch}, is universally recognized as the key resource in the processing of quantum information and an important tool for the implementation of quantum communication and quantum-empowered metrology~\cite{hpower4}. Yet, entanglement does not embody the {\it unique} way in which non-classical correlations can be set among the elements of a composite system. When generic mixed states are considered, quantum correlations (QCs) are no longer synonymous of entanglement: {\it Other} forms of stronger-than-classical correlations exist and can indeed be enforced in the state of a multipartite mixed system. However, a general consensus on {\it the} measure of quantum correlations is still far from having been found. Among the quantifiers proposed so far, quantum discord~\cite{QD} (${\cal D}$)  occupies a prominent position and enjoys a growing popularity within the community working on quantum information science due to its alleged relevance in the model for deterministic
quantum computation with one qubit~\cite{datta,barbieri}, extendibility to some important classes of infinite-dimensional systems~\cite{parisadesso} and peculiar role in open-system dynamics~\cite{laura}. Recently, some attempts at providing an operational interpretation to discord have been reported~\cite{opInt}. 

Yet, interesting alternative to discord exist, each striving at capturing different facets of QCs~\cite{others}. In Ref.~\cite{giro10qph}, in particular, a measure based on the concept of perturbation on a bipartite quantum state (see also Luo in Ref.~\cite{others}) induced by joint local measurements has been put forward and extensively analyzed. Such indicator, dubbed {\it ameliorated measurement induced disturbance} (AMID), has been shown to signal faithfully fully classical states ({\it i.e.} states endowed with only classical correlations). AMID embodies an interesting upper bound to the non-classicality content quantified by {\qd} and, at variance with the latter, is naturally symmetric.
\begin{figure}
\centering
\includegraphics[width=7cm]{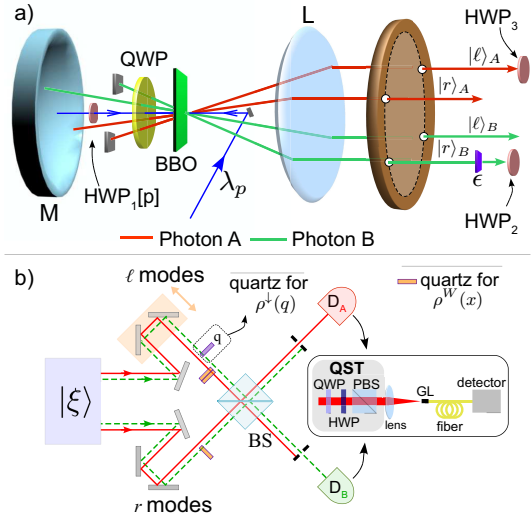}
\caption{(color online) {\bf a)}: Setup for the generation of a polarization-path 4-qubit entangled state. A Type-I nonlinear $\beta$-barium borate crystal (BBO) is pumped by a vertically polarized UV laser at wavelength $\lambda_p$ in a double-pass configuration. This produces non-maximally entangled polarization states that, through the quarter-waveplate QWP, the half-waveplate HWP$_1$ and mirror M, can be turned into the state {$\ket{\phi^+(p)}_{AB}$} described in the body of the paper. The four-hole mask selects four longitudinal spatial modes within the emission cone of the BBO crystal. The attenuator $\epsilon$ and the half-waveplates HWP$_{2,3}$ allow us to engineer the 4-qubit polarization-path entangled state $\ket{\xi}_{AB}$. {\bf b)}: Interferometer needed to perform the trace over the  path DOF and generate the states studied in our investigation (BS stands for a beam splitter). Quartz plates of various thickness have been used to produce $\rho^{\downarrow}(q)_{AB}$ and $\rho^W(\epsilon)_{AB}$. 
We also show the apparatuses $D_{j}~(j{=}A,B)$ needed to perform the quantum state tomography of the states thus generated. Each $D_j$ is made of an analyzer formed by the cascade of a QWP, an HWP and a polarizing beam splitter (PBS). The signal then enters a photodetector. 
}
\label{setup}
\end{figure}

A landmark on the study of quantum entanglement has been set by the identification of states maximizing the degree of two-qubit entanglement at set values of the global state mixedness~\cite{mems}. This has spurred an extensive investigation, at all levels, on the interplay between entanglement and mixedness, which has culminated in the experimental exploration of the two-qubit entropic plane, including maximally entangled mixed states (MEMS) by a number of groups worldwide~\cite{pete04prl,barb04prl,woerdman}. Needless to say, given the strong interplay between non-classical correlations and mixedness,  an experimental characterization analogous to the one performed for MEMS is not only highly desirable but extremely interesting. This is precisely the aim of this work: Building on the framework provided by the theoretical studies in Refs.~\cite{giro10qph,jamesmaiorca}, here we experimentally navigate the space of two-qubit discorded states focusing our attention, in particular, on the class of  two-qubit maximally non-classical mixed states (MNCMS), {\it i.e.} those states maximizing the degree of quantum discord at assigned values of their global von Neumann entropy. We show a very good agreement between theoretical predictions and experimental evidence across the whole range of values of the global entropy for two-qubit states. The extensive nature of our investigation comprises the generation and analysis of a variety of quantum correlated two-qubit states, from Werner states to the MEMS associated with the use of relative entropy of entanglement and von Neumann entropy~\cite{mems}.  

Technically, this has been possible due to the {high} flexibility of the experimental setup used for our demonstration, which makes clever and effective use of the possibilities offered by well-tested sources for hyperentangled polarization-path photonic states. We engineer mixedness in the joint polarization state of two photonic qubits by tracing out the path degree of freedom (DOF). The properties of such residual states are then analyzed by means of the quantum state tomography (QST) toolbox~\cite{jame01pra} and a quantitative comparison between their quantum-correlation contents and the predictions on MNCMS is performed. The quality of the generated states is such that we have been able to experimentally verify the predictions given in Ref.~\cite{giro10qph} relating discord and {AMID}: We have generated the states embodying both the lower and upper bound to {AMID} at set values  of discord.  Our study should be regarded as the counterpart, dealing with the much broader context of general quantum correlations, of the seminal experimental investigations on the relation between entanglement and mixedness performed in Refs.~\cite{pete04prl,barb04prl,woerdman}. As such, it encompasses an important step in the characterization of non-classicality in general two-qubit states. 

\noindent
{\it Resource-state generation.--} Before exploring the entropic two-qubit space, it is convenient to introduce the experimental techniques used in order to achieve the ample variety of states necessary for our investigation. The key element for the state engineering in our setup is embodied by state 
\begin{equation}\label{xi}
\ket{\xi}_{AB}{=}\sqrt{1{-}\epsilon}\ket{r\ell}_{AB}\ket{\phi^+(p)}_{AB}+\sqrt{\epsilon} \ket{\ell r}_{AB}\ket{HV}_{AB}
\end{equation}
with {$\ket{\phi^\pm(p)}_{AB}{=}\sqrt{p}\ket{HH}_{AB}{\pm}\sqrt{1-p}\ket{VV}_{AB}$}. 
In Eq.~(\ref{xi}), four qubits are encoded in the polarization and path DOFs
 of optical modes $A$ and $B$. In particular, $H$ ($V$) represents horizontal (vertical) polarization of a photon, while $r$ ($\ell$) 
is the right (left) mode in  which each photon can be emitted from our source of entangled photon states, which we now describe. 
State $\ket{\xi}_{AB}$ is produced by suitably adapting the polarization-momentum source of hyperentangled states that has been recently used as basic building block in experimental test-beds on multipartite entanglement~\cite{barb05pra, chiu10prl}.
{To generate $\ket{\phi^{+}(p)}$, 
a UV laser impinges back and forth on a nonlinear crystal [cfr. Fig.~\ref{setup} {\bf a)}]. 
The forward emission generates the $\ket{HH}$ contribution.
A quarter waveplate (QWP) transforms the $\ket{HH}$ backward 
emission into $\ket{VV}$ after reflection at the spherical mirror $M$.}
The relative phase between the $\ket{VV}$ and $\ket{HH}$ contributions is changed by
translating $M$. The weight $\sqrt p$ in the unbalanced Bell state 
{$\ket{\phi^{+}(p)}_{AB}$}, 
can be varied by rotating the
half waveplate HWP$_{1}[p]$ near $M$ [see Fig.~\ref{setup} {\bf a)}], which intercepts twice the UV pump beam. For more 
details on the generation of non-maximally entangled states of polarization, see Ref.~\cite{vall07pra}. 
A four-hole mask allows us to select four longitudinal spatial modes (two per photon), 
namely $\ket{r}_{A,B}$, $\ket{\ell}_{A,B}$,  
within the emission cone of the crystal.
The state thus produced finally reads $\ket{\text{HE}(p)}{=}(\ket{r\ell}_{AB}+e^{i\gamma}\ket{\ell r}_{AB}) \otimes \ket{\phi^{+}(p)}_{AB}/{\sqrt{2}}$.

State $\ket{\xi}_{AB}$ has been obtained by making three further changes to
$\ket{\text{HE}(p)}$ [cfr. Fig.~\ref{setup} {\bf a)}]. First, the contributions of modes $\ket{\ell r}$ corresponding to the V-cone is intercepted by inserting two beam stops. An attenuator is then
placed on mode $\ket{r}_{B}$ so as to vary 
 the relative weight between  $\ket{\ell r}_{AB}$ and $\ket{r \ell}_{AB}$. This effectively corresponds to changing $\epsilon$. Finally, a HWP [labelled HWP$_{2}$ in Fig.~\ref{setup} {\bf a)}], oriented at 45$^{\circ}$ and intercepting mode $\ket{r}_{B}$, 
 allows to transform $\ket{\ell r}\ket{HH}$ into $\ket{\ell r}\ket{HV}$. This gives us the  second term in \eq{xi}, with which we have been able to span the entire set of states relevant to our study. 
\begin{figure*}[t]
\includegraphics[width=13cm]{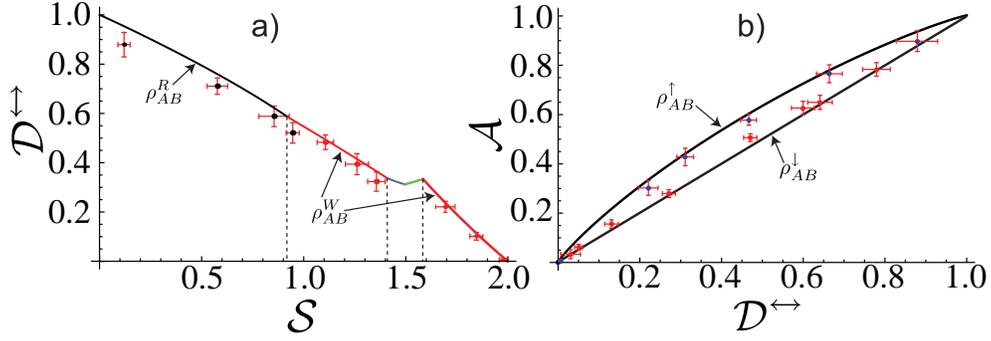}
\caption{(color online) {\bf a)}: Exploration of the ${\cal D}^{\leftrightarrow}\ vs. \ {\cal S}$ plane. The multi-color solid line shows the MNCMS boundary. 
{\bf b)}: Experimental comparison between AMID and ${\cal D}^\leftrightarrow$. The upper and lower solid lines embody the bounds to ${\cal A}$ at set value of symmetrized discord. Both panels show experimental states and associated uncertainties.}
\label{disc-vne&amid-disc}
\end{figure*}

\noindent
{\it Experimental navigation.--} We now introduce the measures of QCs considered in our work and discuss the results of our experimental investigation. We start reminding that discord is associated to the discrepancy between two classically
equivalent versions of mutual information~\cite{QD}. For a bipartite state $\rho_{AB}$ 
the  latter is defined as ${\cal I}(\rho_{AB}){=}{\cal S}(\rho_A){+}{\cal
S}(\rho_B){-}{\cal S}(\rho_{AB})$. Here, ${\cal
S}(\rho){=}{-}\text{Tr}[\rho\log_2\rho]$ is the von Neumann entropy (VNE)
of the arbitrary two-qubit state $\rho$ and $\rho_{j}$ is the reduced
density matrix of party $j{=}A,B$. One can also consider the expression
${\cal J}^\leftarrow(\rho_{AB}){=}{\cal S}(\rho_A){-}{\cal
H}_{\{\hat\Pi_i\}}(A|B)$ (the one-way classical correlation~\cite{QD}) with ${\cal
H}_{\{\hat\Pi_i\}}(A|B){\equiv}\sum_{i}p_i{\cal S}(\rho^i_{A|B})$ the quantum
conditional entropy associated with the the post-measurement density matrix
$\rho^i_{A|B}{=}\text{Tr}_B[\hat\Pi_i\rho_{AB}]/p_i$ obtained upon
performing the complete projective measurement $\{\Pi_i\}$ on system $B$
($p_i{=}\text{Tr}[\hat\Pi_i\rho_{AB}]$). We define discord as 
${\cal D}^\leftarrow{=}\inf_{\{\Pi_i\}}[{\cal
I}(\rho_{AB}){-}{\cal J}^\leftarrow(\rho_{AB})]$,
where the
infimum is calculated over the set of projectors $\{\hat\Pi_i\}$.
Discord is in general asymmetric (${\cal D}^\leftarrow{\neq}{\cal
D}^\rightarrow$) with ${\cal D}^\rightarrow$ obtained by swapping the roles of A
and B. This originates the possibility to distinguish between {\it quantum-quantum states} having $({\cal D}^\leftarrow,{\cal D}^\rightarrow){\neq}{0}$, {\it quantum-classical} and {\it classical-quantum} ones, which are states having one of the two values of discord strictly null, and finally {\it classical-classical} states for which ${\cal D}^\leftarrow,{\cal D}^\rightarrow{=}{0}$, which are bipartite states that simply embed a classical probability distribution in a two-qubit state~\cite{pianista}. Clearly, the asymmetry inherent in discord would lead us to mistake a quantum-classical state 
as a classical state. This makes such a measure not strongly faithful. In order to bypass such an ambiguity we will consider the symmetrized discord ${\cal D}^\leftrightarrow{=}\max[{\cal D}^\leftarrow,{\cal D}^\rightarrow]$, which is zero only for classical-classical states. 

In Ref.~\cite{giro10qph}, AMID has been introduced as an alternative indicator of non-classical correlations for bipartite systems of any dimension as
${\cal A}
{=}{\cal I}(\varrho_{AB}){-}{\cal I}_c(\varrho_{AB})$,
where ${\cal I}_c(\varrho_{AB}) \equiv {\sup}_{\{\hat\Omega\}}{\cal
I}(\varrho^{\hat{\Omega}}_{AB})$ and $\varrho^{\hat\Omega}_{AB}$ is the state resulting from the application of the arbitrary complete (bi-local) projective measurement over the composite system $\hat{\Omega}_{kl}=\hat\Pi_{A,k}\otimes\hat\Pi_{B,l}$.
Our definition is motivated by the analysis in~\cite{terhal}, where ${\cal I}_c$ is defined as the {\it classical mutual information} (optimized over projective measurements), a proper
symmetric measure of classical correlations in bipartite states.
 AMID is thus  recast as the difference between total and classical mutual
information, which has all the prerequisites to be a {\it bona fide} measure
of QCs \cite{pianista}.

Having presented our quantitative tools, we are in a position to discuss the results of our experimental endeavors by first addressing the ${\cal D}^\leftrightarrow\ vs.\ {\cal S}$ plane. As shown in Refs.~\cite{giro10qph} (see also~\cite{jamesmaiorca}), when ${\cal D}^\leftrightarrow$ and ${\cal S}$ are taken as quantitative figures of merit for QCs and global mixedness respectively, the class of MNCMS consists of four families of states, all of the form
\beq
\rho_{AB}^{X} = \left[
\begin{matrix}
\rho_{11} & 0 & 0 & \rho_{14}\\
0 & \rho_{22} & \rho_{23} & 0\\
0 & \rho_{23}^{*} & \rho_{33} & 0\\
\rho_{14}^{*} & 0 & 0 & \rho_{44}
\end{matrix} 
\right]\!,~~\text{with}~~\sum_{j}\rho_{jj}{=}1.
\eeq
The low-entropy region ${\cal S}{\in}[0, 0.9231)$ pertains to the rank-3 states $\rho^R_{AB}$ embodying MEMS for the relative entropy of entanglement~\cite{mems}
\beq
\label{cc}
{\rho^R_{AB}=\frac{1-a+r}{2}\ket{\Phi^+}\bra {\Phi^+}+
\frac{1-a-r}{2}\ket{\Phi^-}\bra {\Phi^-}+a\ket{01}\bra{01}}
\eeq
with $0{\le}a{\le}1/3$ and $r$ a proper function of $a$~\cite{giro10qph}. 
{In Eq.~(\ref{cc}) we have used the Bell state $\ket{\Phi^\pm}\equiv\ket{\phi^{\pm}(1/2)}_{AB}$}. 
States $\rho^R_{AB}$ span the black-colored trait in Fig.~\ref{disc-vne&amid-disc} {\bf a)}. Next comes the family of Werner states 
\beq
{\rho^{W}_{AB}(\epsilon){=}(1{-}\epsilon) 
\ket{\Phi^{+}}_{AB} \bra{\Phi^{+}}{+} \epsilon \frac{\id_{4}}{4}}, 
\eeq
which occupy the entropic sector ${\cal S}{\in}[0.9231, 1.410)$ for {$0.225{\le}\epsilon{<}{0.426}$} 
and the high-entropy region of ${\cal S}{\in}[1.585, 2]$ for {${0.519}{\le}{\epsilon}{\le}{1}$}. 
Such disjoint boundaries are both represented by red curves in Fig.~\ref{disc-vne&amid-disc} {\bf a)}. 
There, it is visible that two more families belong to the MNCMS boundary [cfr. the blue and green traits corresponding to the range $1.410{\le}{\cal S}{<}1.585)$]. Such states are currently out of our grasp due to the rather small entropy-window they belong to, which poses some challenge to the tunability of the state mixedness achievable by our method. For conciseness, we omit any discussion about them. 


It is worth noticing that quantum discord and AMID share the very same structure of MNCMS, which can thus be rightfully regarded as the two-qubit states whose QCs are maximally robust against state mixedness. This class of states are thus set to play a key role in realistic (noisy) implementations of quantum information schemes based on non-classicality of correlations as a resource~\cite{datta,barbieri}. There is, currently, an enormous interest in designing practical schemes for the exploitation of such features. The sharing of such interesting class of states by the two QC measures addressed here enforces the establishment of a hierarchy between AMID and quantum discord, a point that is precisely along the lines of interesting quantitative comparisons between different measures of entanglement applied to mixed two-qubit states~\cite{grudka}, in an attempt to establish a mutual order.

Such a relationship is elucidated in Fig.~\ref{disc-vne&amid-disc} {\bf b)}, where the solid lines show that AMID embodies an upper bound to ${\cal D}^\leftrightarrow$ and is in agreement with the latter in identifying genuinely classical-classical states having no QCs. Any physically allowed two-qubit state lives in between the straight lower bound such that ${\cal A}{=}{\cal D}^\leftrightarrow$ and the upper one. A full analytic characterization of such boundary curves is possible and can be thoroughly checked by means of a numerical exploration of the ${\cal A}\ vs.\ {\cal D}^\leftrightarrow$ plane~\cite{giro10qph}. Quite obviously, the lower bound in the AMID-discord plane is spanned by pure states of variable entanglement (for pure states ${\cal A}{=}{\cal D}^\leftrightarrow$).
However, such a lower frontier also accommodates both the Werner states 
and the family
\beq\label{low}
{\rho^{\downarrow}_{AB}(q)=(1-q) \ket{\Phi^{+}}_{AB}\bra{\Phi^{+}}+ q \ket{\Phi^{-}}_{AB}\bra{\Phi^{-}}},
\eeq
where $q{\in}[0,0.5]$, while  the upper bound is spanned by
\beq\label{up}
{\rho^{\uparrow}(\epsilon,p)_{AB}{=}(1{-}\epsilon)\ket{\phi^{+}(p)}_{AB}\bra{\phi^{+}(p)}{+}\epsilon \ket{01}_{AB} \bra{01}}
\eeq
for values of $(\epsilon,p)$ satisfying a transcendental equation~\cite{giro10qph}.
Starting from the {four-qubit} resource $\ket{\xi}_{AB}$, we have generated the states spanning the MNCMS boundary in Fig.~\ref{disc-vne&amid-disc} {\bf a)} and the upper/lower frontier states $\rho^{\updownarrow}_{AB}(\epsilon,p)$ in the ${\cal A}\ vs\ \mathcal{D}^{\leftrightarrow}$ plane. 

\noindent
{{\it Generation of $\rho^{\uparrow}_{AB}$.-} This class} serves an ideal platform for the description of the experimental method pursued to achieve the remaining states addressed in our study. By tracing out the path DOF {in $\ket{\xi}_{AB}$}
and using the correspondence between physical states and logical qubits $\ket{H}{\rightarrow}\ket{0}$, $\ket{V}{\rightarrow}\ket{1}$,
the density matrix for state $\rho^{\uparrow}(\epsilon,p)$ is achieved.
The trace over the path DOF is performed by matching the left and right side of the 
modes coming from the four-hole mask in Fig.~\ref{setup} {\bf a)} on a beam splitter [indicated as BS in panel {\bf b)} of the same figure]. 
When the difference between left and right paths 
is larger than the photon coherence time,
an incoherent superposition of {$\ket{\phi^{+}(p)}_{AB}$} and $\ket{HV}_{AB}$
is  achieved. The values of the pairs $(\epsilon,p)$ determining the experimental
states [shown as blue dots in Fig.~\ref{disc-vne&amid-disc} {\bf b)}] are given in Table~\ref{value}, together with their uncertainties.
\begin{table}[t]
  \begin{tabular}{ c|  c  c  c  c  c  c }
    \hline
    \hline
    & & Value & and & uncertainty & \\
    \hline\hline
    $\epsilon$&0.00$\pm$0.01\!&0.05$\pm$0.01\!& 0.10$\pm$0.01\!& 0.15$\pm$0.01\!& 0.18$\pm$0.01\!&0.20$\pm$0.01 \\ 
    $p$&0.50$\pm$0.02\!&  0.70$\pm$0.01\!&  0.80$\pm$0.01\!&  0.90$\pm$0.01\!&0.95$\pm$0.02\!&0.99$\pm$0.02 \\ 
    \hline
    \hline
  \end{tabular}
  \caption{{\bf Parameters in $\rho^{\uparrow}(\epsilon,p)$}: The table reports the values of the parameters entering the states $\rho^{\uparrow}(\epsilon,p)$ produced in our experiment, together with their uncertainties.}
\label{value}
\end{table}
The values ($\epsilon,p$)=(0,0.5) and ($\epsilon,p$)=(0.2,1) correspond to the case of a pure state (having $\mathcal{A}{=}\mathcal{D}^{\leftrightarrow}{=}1$)
and a completely mixed state (with $\mathcal{A}{=}\mathcal{D}^{\leftrightarrow}${=}0) respectively. 

\noindent
{\it Generation of $\rho^{\downarrow}_{AB}$.- }
The family embodied by $\rho^{\downarrow}_{AB}(q)$ can also be generated starting from the resource state $\ket{\xi}_{AB}$. By selecting only the correlated modes $\ket{r\ell}_{AB}$ from the four-hole mask and setting the HWP$_{1}$ at $0^\circ$ (so that $p=1/2$ is fixed), we have generated the Bell state 
{$\ket{\Phi^{+}}_{AB}\bra{\Phi^{+}}{\equiv}\rho^{\downarrow}(q{=}0)$.} 
By inserting a birefringent quartz plate of proper thickness on the path of one of the two correlated modes, we controllably affect 
{the coherence between the $\ket{HH}$ and $\ket{VV}$} states of polarization. 
Several quartz plates of different thickness $\ell_{q}$ have been used 
{to transform $\ket{\Phi^+}$ into $\rho^{\downarrow}(q)$. 
The} value of $q$ is related to the dimensionless parameter $C=\frac{(\Delta n)\ell_{q}}{c\tau_{coh}}$,
where $\tau_{coh}$ is the coherence time of the emitted photons and $\Delta n$ is the difference between ordinary and extraordinary
refraction indices in the quartz. The details of such dependence are inessential and it is enough to state that $q{=}1/2$ ($q\rightarrow0$) for $C{\gg}1$ ($C{\rightarrow}0$).

\noindent
{{\it Generation of $\rho^{R,W}$.-} 
Our source of $\rho^R$ and Werner state makes use of the setup previously described for the 
states $\rho^{\uparrow}_{AB}(\epsilon,p)$. 
By setting $p=1/2$ and by adding a decoherence between $\ket{HH}$
and $\ket{VV}$ (related to the parameter $r$) as previously explained, we can obtain $\rho^R_{AB}$ from $\rho^{\uparrow}_{AB}$. } 
As for $\rho^W_{AB}$, while we have already addressed the method used to generate the {$\ket{\Phi^{+}}_{AB}\bra{\Phi^{+}}$}  component 
of the state, it is worth mentioning how to get the $\id_{4}$ contribution. 
This has been obtained by inserting a further HWP [HWP$_{3}$ in Fig.\ref{setup} {\bf a)}]
on the $\ket{\ell}_{A}$ mode and rotating both HWP$_{2}$ and HWP$_{3}$ at $22.5^{\circ}$ so as to generate $\ket{\ell r}_{AB} \ket{++}_{AB}$. By using two quartz plates longer than $\tau_{coh}$
and of different thickness, we obtained a fully mixed state on the correlated modes $\ket{\ell r}_{AB}$. Each quartz plate introduces decoherence on the state of each photon. By matching the two correlated-mode pairs on a BS, state $\rho^W_{AB}$ is achieved.


As anticipated, in order to ascertain the properties of all the states being discussed above, we have used QST~\cite{jame01pra} so as to obtain the corresponding physical density matrices and quantify $\mathcal{D}^{\leftrightarrow}$, $\mathcal{S}$ and $\mathcal{A}$. The Pauli operators needed to implement the QST have been measured by using standard polarization analysis setup and two detectors [see the inset in Fig.\ref{setup} {\bf b)}]. Integrated systems given by GRIN lenses and single mode fibres~\cite{ross09prl} have been used to optimally collect the radiation after the QST setup and send it to the detectors $\text{D}_{A,B}$.

\noindent
{\it Discussion and conclusions.--} 
{Excellent agreement between the theoretical expectations and experimental results has been found}
for both the navigation in the space of MNCMS and the quantitative confirmation of the predicted relation between AMID and discord. As seen in Fig.~\ref{disc-vne&amid-disc}, almost the whole class of maximally non-classical states has been explored, with the exception of a technically demanding (yet interesting) region, whose exploration is currently under study. Quite remarkably, on the other hand, the whole upper bound in the ${\cal A}\ vs. \ {\cal D}^\leftrightarrow$ has been scanned in an experimental endeavor that has originated an ample wealth of physically interesting states. 
{Technically, this has been achieved by cleverly engineering a four-qubit hyperentangled state.
In particular, we exploits the path as an ancillary resource to obtain the desired states 
encoded in the polarization. } Our analysis remarkably embodies the first navigation in the space of general quantum correlations at set values of global entropy, thus moving along the lines of the analogous seminal investigations performed on entanglement~\cite{pete04prl,barb04prl,woerdman}. We {hope} that our efforts will spur further interest in the study, at all levels, of the interplay between  mixedness and non-classicality.



\noindent
{\it Acknowledgments.--} 
{We thank Valentina Rosati for the contributions given in realizing the experiment. MP is grateful to G. Adesso and D. Girolami for fruitful discussions. }
{This work was partially supported by the FARI project 2010 of Sapienza Universit\`a di Roma,}
and the UK EPSRC (EP/G004579/1).



\begin{thebibliography}{10}
\providecommand{\url}[1]{\texttt{#1}}
\providecommand{\urlprefix}{URL }
\providecommand{\eprint}[2][]{\url{#2}}

\bibitem{Sch} E. Schr\"odinger, Proc. Cambridge Philos. Soc. {\bf 31}, 555 (1935).

\bibitem{hpower4} R. Horodecki {\it et al.}, Rev. Mod. Phys. {\bf 81}, 865 (2009).

\bibitem{QD} H. Ollivier and W. H. Zurek, Phys. Rev. Lett. {\bf 88}, 017901 (2001); L. Henderson and V. Vedral, J. Phys. A {\bf 34}, 6899 (2001).

\bibitem{datta} A. Datta, {\it et al.}, Phys. Rev. Lett. {\bf 100}, 050502 (2008).

\bibitem{barbieri} B. P. Lanyon, {\it et al.}, Phys. Rev. Lett. {\bf 101}, 200501 (2008).

\bibitem{parisadesso} P. Giorda, and M. G. A. Paris, Phys. Rev. Lett. {\bf 105}, 020503 (2010); G. Adesso and A. Datta, {\it ibid.} {\bf 105}, 030501(2010).

\bibitem{laura} L. Mazzola, {\it et al.}, Phys. Rev. Lett. {\bf 104}, 200401 (2010).

\bibitem{opInt} D. Cavalcanti, {\it et al.}, arXiv:1008.3205; V. Madhok, and A. Datta, arXiv:1008.4135.

\bibitem{others} S. Luo, Phys. Rev. A {\bf 77}, 042303 (2008). 

\bibitem{giro10qph}
D.~Girolami, M.~Paternostro, and G.~Adesso, 
  \eprint{arXiv:1008.4136}.

\bibitem{mems} W. J. Munro, {\it et al.}, Phys. Rev. A {\bf 64}, 030302 (2001); T. Wei, {\it et al.}, Phys. Rev. A {\bf 67}, 022110 (2003).

\bibitem{pete04prl}
N.~A. Peters, {\it et al.},
Phys. Rev. Lett. \textbf{92}, 133601 (2004).

\bibitem{barb04prl}
M.~Barbieri, {\it et al.},
Phys. Rev. Lett.
  \textbf{92}, 177901 (2004).

\bibitem{woerdman} G. Puentes, {\it et al.}, Phys. rev. A {\bf 75}, 032319 (2007).

\bibitem{jamesmaiorca} A. Al-Qasimi, and D. F. V. James, Phys. Rev. A {\bf 83}, 032101 (2011); F. Galve, {\it et al.}, {\it ibid.} {\bf 83}, 012102 (2011).
\bibitem{jame01pra}
D.~F.~V. James, {\it et al.},
Phys. Rev. A  \textbf{64}, 052312 (2001).

\bibitem{chiu10prl}
A.~Chiuri, {\it et al.},
  Phys. Rev. Lett. \textbf{105}, 250501 (2010).

\bibitem{vall07pra} 
G. Vallone, {\it et al.}, Phys. Rev. A {\bf 76}, 012319 (2007).

\bibitem{pianista}  
M. Piani, P. Horodecki, and R. Horodecki,  {Phys. Rev. Lett.} {\bf 100}, 090502 (2008).

\bibitem{terhal} 
B. M. Terhal, {\it et al.},  J. Math. Phys. {\bf 43}, 4286 (2002); 
D. P. DiVincenzo, {\it et al.}, Phys.
Rev. Lett. {\bf 92}, 067902 (2004).


\bibitem{barb05pra}
M.~Barbieri, {\it et al.},
Phys. Rev. A
  \textbf{72}, 052110 (2005).

\bibitem{zhan02pra}
Y.-S. Zhang, {\it et al.},
Phys. Rev. A \textbf{66},
  062315 (2002).

\bibitem{ross09prl}
A.~Rossi, {\it et al.},
Phys. Rev.
  Lett. \textbf{102}, 153902 (2009).


\bibitem{grudka} S. Virmani, and M. B. Plenio, Phys. Lett. A {\bf 268}, 31 (2000); A. Miranowicz, and A. Grudka, J. Opt. B: Quantum Semiclass. Opt. {\bf 6}, 542 (2004); Phys. Rev. A {\bf 70}, 032326 (2004).

\end{thebibliography}

\end{document}